\documentclass[twocolumn,conference]{IEEEtran}
\usepackage[T1]{fontenc}
\usepackage[latin9]{inputenc}
\usepackage{array}
\usepackage{varioref}
\usepackage{calc}
\usepackage{textcomp}
\usepackage{multirow}
\usepackage{amstext}
\usepackage{graphicx}
\usepackage[unicode=true,
 bookmarks=true,bookmarksnumbered=true,bookmarksopen=true,bookmarksopenlevel=1,
 breaklinks=false,pdfborder={0 0 0},pdfborderstyle={},backref=false,colorlinks=false]
 {hyperref}
\hypersetup{pdftitle={Your Title},
 pdfauthor={Your Name},
 pdfpagelayout=OneColumn, pdfnewwindow=true, pdfstartview=XYZ, plainpages=false}

\makeatletter

\providecommand{\tabularnewline}{\\}

\usepackage[caption=false,font=footnotesize]{subfig}
\renewcommand{\textendash}{--}

\newif\ifanonym
\anonymfalse
\newif\ifbiblioinside
\biblioinsidetrue

\makeatother

\begin{document}

\title{Table understanding in structured documents}

\author{\ifanonym\else
\IEEEauthorblockN{Martin~Hole\v{c}ek\IEEEauthorrefmark{1}\IEEEauthorrefmark{2}, Antonín~Hoskovec\IEEEauthorrefmark{2},
Petr~Baudiš\IEEEauthorrefmark{2}, Pavel~Klinger\IEEEauthorrefmark{2}}\IEEEauthorblockA{\IEEEauthorrefmark{1}Faculty of Mathematics and Physics, Charles
University, Department of Numerical Mathematics}\IEEEauthorblockA{\IEEEauthorrefmark{2}Rossum}\fi}
\maketitle
\begin{abstract}
Table detection and extraction has been studied in the context of
documents like reports, where tables are clearly outlined and stand
out from the document structure visually. We study this topic in a
rather more challenging domain of layout-heavy business documents,
particularly invoices. Invoices present the novel challenges of tables
being often without outlines - either in the form of borders or surrounding
text flow - with ragged columns and widely varying data content. We
will also show, that we can extract specific information from structurally
different tables or table-like structures with one model. We present
a comprehensive representation of a page using graph over word boxes,
positional embeddings, trainable textual features and rephrase the
table detection as a text box labeling problem. We will work on our
newly presented dataset of pro forma invoices, invoices and debit
note documents using this representation and propose multiple baselines
to solve this labeling problem. We then propose a novel neural network
model that achieves strong, practical results on the presented dataset
and analyze the model performance and effects of graph convolutions
and self-attention in detail.
\end{abstract}

\begin{IEEEkeywords}
table detection; neural networks; invoices; graph convolution; attention
\end{IEEEkeywords}

\IEEEpeerreviewmaketitle{}

\section{Introduction}

Table detection and table extraction problems were already introduced
in a competition ICDAR 2013, where the goal was to detect tables and
extract cell structures from a dataset of mostly scientific documents
\cite{gobel2013icdar}. Table can be defined as a set of content cells
organized in a self-describing manner into such a structure, that
groups cells into rows and columns. This was reflected in the metric
defined in the competition, that scores tables based on the relations
successfully extracted. Similarly, structured documents do have a
self-describing structure, that often looks table-like.

We have decided to investigate the problem on business documents such
as invoices, pro forma invoices and debit notes (referred for simplicity
as invoices or invoice-type documents later in this text), where the
aim is different. Namely - even table detection needs to be thought
of in the context of document understanding, because invoices are
inherently documents with textual information structured into more
tables. Graphical borders and edges are sometimes present, however,
they cannot be used for detection, because there is no general layout
and very often there are no borders at all. Another obstacle for traditional
methods is the fact, that the data can span over multiple lines of
text which holds true also for the table cells.

Moreover, we require our model to 'understand' the document in a way
that it could classify tables and tabular structures based on their
content. In practice the goal is to detect the whole table with the
so-called 'line-items' (detailed items of the total amount to pay)
and, at the same time, extract only a specific information from the
other tables (to find a 'field'). Simply said, not every table inside
an invoice should be detected and reported as whole (see example invoice
on figure \vref{fig:Sample-invoice-edges}). Usually in commercial
applications this problem is tackled using a layout system that detects
the layout and extracts the table (or a field) from a position where
it usually happens to be; or employing another classification module,
which selects the right table from several proposals. That increases
the number of modules in the architecture and requires manual layout
setups, while our goal is to have a trainable system that could leverage
the commonalities present in the data without ongoing human support.
To verify that, we will ensure that proper generalization of models
predictions is evaluated on new layouts.

\section{Previous Work}

The plethora of methods that have been previously used for the task
is hard to summarize or compare since all the algorithms have been
used/evaluated on different datasets and each have their strengths,
weak spots and quirks. However, we found none of them well suited
for working with structured documents (like invoices), since they
in general have no fixed layout, language, captions, delimiters, fonts...
For example, invoices vary in countries, companies and departments
and changes in time. In order to retrieve any information from a structured
document, you need to understand it.

In literature there are examples of table detection using heuristics
\cite{2014arXiv1412.7689J}, using layouts \cite{dhiran2013table},
regular expressions \cite{mandal2004very}, or leveraging the presence
of lines in tables \cite{gatos2005automatic,tablemetadata,liu2009tableseer,interpretingscanned},
or using clustering \cite{pether2016robust}. A great survey can be
found in \cite{milovsevic2018table}.

Tables were searched for also in HTML \cite{tengli2004learning,Chu:2015:TTE:2723372.2723725},
free text \cite{Ng:1999:LRT:1034678.1034746} or scientific articles
with a method based on matching captions with content \cite{clark2015looking}.

Machine learning methods and deep neural networks were also employed
in several papers. The work \cite{2015arXiv150608891F} aims at scientific
documents using fine tailored methods stacked atop each other. Reference
\cite{detectdeep} uses Fast R-CNN architecture with a novel idea
of Euclidean distance feature to detect tables (which was compared
to Tesseract). Reference \cite{8270123} also uses (pretrained) Fast
R-CNN and FCN semantic segmentation model for table extraction problem.
In \cite{interpretingscanned} work has been done on detection problem
bottom up using the Hough transform, and extraction was solved with
Markov networks and features from the cell positions. Reference \cite{10.1007/978-3-540-24580-3_54}
uses convolutions over the number (and sizes) of spaces in a line.
A deep CNN approach was being investigated in \cite{Kavasidis2018ASC},
which combined CNNs for detecting row and columns, dilated convolutions,
CRFs and saliency maps, they have also developed a webcrawler to extend
their dataset. We tried and failed to get working results using the
YOLO architecture \cite{yolov3} with textual datasets. (We have experimented
with YOLO because some works aimed at table detection do use the family
of R-CNNs, Fast R-CNNs and Mask R-CNNs, that preceded the development
of YOLO.)

For document understanding, a graph representation of a document was
examined in \cite{8395204,Couasnon2014}, finding similar documents
and reusing their goldstandards was done in \cite{10.1007/978-3-540-74141-1_28}.

\section{Methodology}

We would like to define our target as creating a model for tabular
or structured data understanding with relevant information detection
and classification. The basic unit of information will be a word in
a document's page with its placement and possibly other features such
as style (see PDF format text data organization \cite{pdfparser}
for example). In this text, we will be calling them simply as \emph{wordboxes}.

With table understanding we mean a joint task of line-item table detection
and information extraction from other tables. The information to be
extracted is defined by the document use-case or semantics, for line-item
table it is the whole table ('table detection' task as defined in
\cite{gobel2013icdar}), while for other structures it is just a specific
infomation ('information extraction' task). No other constraints apply,
i.e. the data can span over multiple lines. So the model is required
to understand a type of table internally and we hope, that the two
tasks will boost the learning process for each other.

With line-item table detection method, we will understand a model,
that could classify each wordbox in a document as being a part of
a line-item content or not (which basically identifies the table itself,
because all line-items tables happen to be well separated, so no instance
segmentation is needed). Same classification approach will be used
for other classes representing other types of content. The classes
are acquired from expert annotations and, as it turns out, we are
dealing with a multilabel problem, i.e. 35 classes in total, examples
being the total amount or recipient address. Also, not every document
contains instances of every class.

\subsection{Metrics and evaluation}

We will observe the scores at validation and test splits, the test
being composed not only of different data, but also of different layouts
and invoice types, thus allowing us to observe the system scalability.
The scores are:
\begin{itemize}
\item $F_{1}$ scores on line-item wordbox classification averaged from
both positive classes and negative classes. \\
At \cite{gobel2013icdar} a content oriented metric was defined for
table detection on character level - each character being either in
the table or out of the table. For us the basic unit is a wordbox,
hence we will define our metric similarily to be the $F_{1}$ score
of table body wordbox class classification.
\item For other classes we will be looking at micro $F_{1}$ scores (only
from positive classes, because the counts of positive samples are
outnumbered by the negative samples - in total, the dataset contains
$1.2\,\%$ positive classes).
\end{itemize}

We chose micro metric aggregation rule, because it gives higher importance
to bigger documents (in the number of wordboxes) which we consider
being more difficult for both human and machine.

We present our research as a novel approach, because referenced papers
or commercial solutions cannot be customized to fit our aim. So we
will compare only against baseline logistic regression over the model
features.

\subsection{The data and their acquisition process}

The data were acquired as a result of work of annotation and review
teams together with automated preprocessing and error-finding algorithms,
that reported errors in nearly $3\,\%$ of the annotation labels.
Classes were annotated in our annotation apps by drawing a rectangle
over the area with the target text. Manual inspection has revealed,
that the annotations can erroneously overlap portions of neighbouring
words, so for ground truth generation we have decided to select only
the wordboxes that are being overlapped by the annotation rectangle
by more than $20\,\%$ of their area.

\paragraph*{Datasets}

We have a dataset with $3554$ PDF invoice files consisting of $4848$
pages in total. The documents are of various vendors, layouts and
languages, annotated with line-item table header and table body together
with other structural information. And we also have a bigger dataset
of $25071$ PDF files of $35880$ pages with just structural information
without line-items (datasets are noted as 'small' and 'big' in the
results).

The documents are standard PDF files, not scanned documents or documents
captured by a digital camera. This decision will not impact the robustness
of our model - given a process to extract bounding boxes and text,
we can use our method in a straightforward manner.

Validation split is chosen to be 1/4 the size. The validation set
measures adaptation, because it could contain similar invoice types
from similar vendors. So in addition, we have created another testing
set of $83$ documents, that have different invoice layouts and types
to those in the training set to measure generalization.

Since this newly compiled dataset was never explored and made accessible
before, we have published an anonymized version of the small dataset,
that contains only the positions and sizes of wordboxes and annotations,
no picture information and no readable text information \textendash{}
only a subset of some textual features. \ifanonym
\else The
dataset is to be found at \href{https://github.com/rossumai/flying-rectangles}{https://github.com/rossumai/flying-rectangles}
\fi

\subsection{Our approach}

We want to operate based on the principle of reflecting the structure
of the data in the model's architecture, as Machine learning algorithms
tend to perform better that way.

What will be the structured information at the input? The number of
wordboxes per page can vary and so we have decided to perceive the
input as an ordered sequence (see below).

In addition we will teach the network to not only detect line-item
table in general, but also to detect a header in the table, because
that could provide a meaningful information - the headers are always
different from the contents.

\begin{figure}[tbh]
\ifanonym\else\includegraphics[width=1\columnwidth]{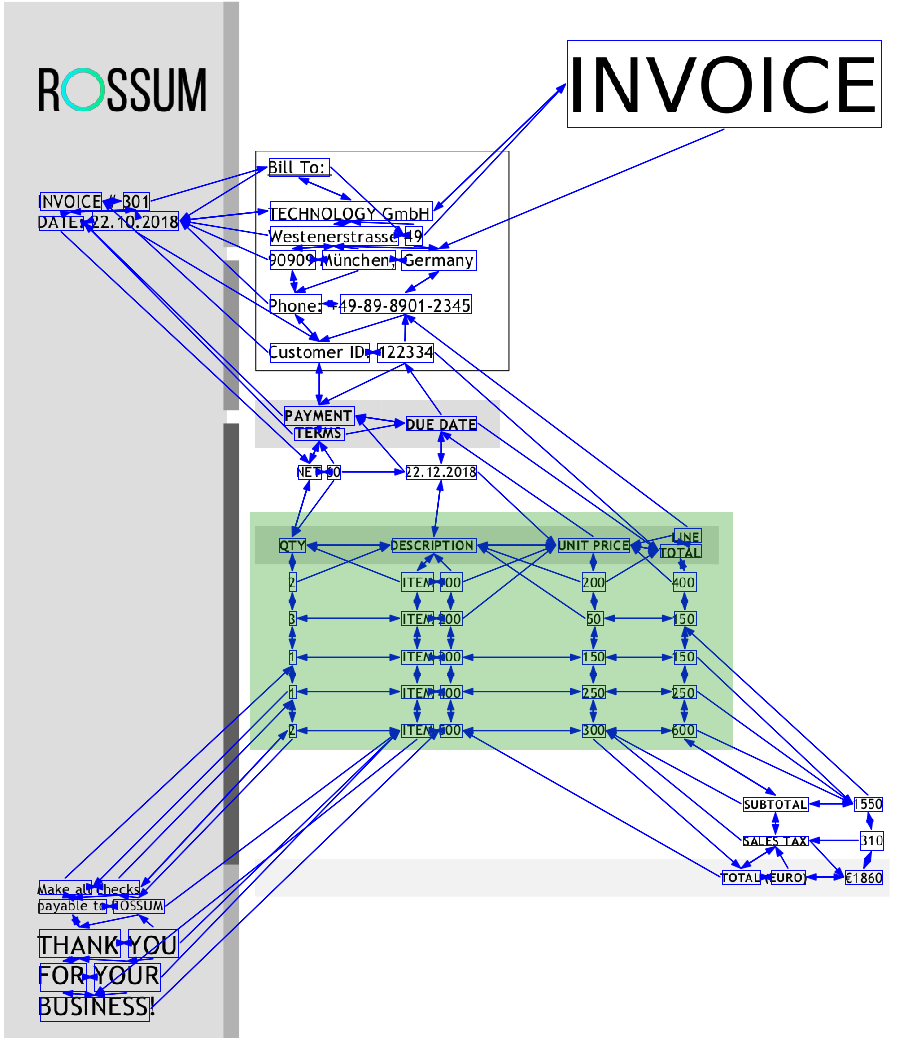}\fi\caption{\label{fig:Sample-invoice-edges}Sample invoice with edges defining
neighbourhood wordboxes. Only the closest neighbour is connected for
each wordbox. Green area is the line-item table. Note that above it
lies a smaller table of payment terms and due date, from which only
some information should be extracted and the table should not be reported.
This invoice is artificially created for presentation and does not
represent the invoices in our dataset.}
\end{figure}

The features of each wordbox are:
\begin{itemize}
\item Geometrical:
\begin{itemize}
\item By geometrical algorithms we can construct a neighbourhood graph over
the boxes, which can then be used by a graph CNN if we bound the number
of neighbours on each edge of the box by a constant. \\
Neighbours are generated for each wordbox ($W$) as follows - every
other box is assigned to an edge of $W$, that has it in its field
of view (being fair $90\text{\textdegree}$), then the closest (center
to center Euclidian distance) $n$ neighbours are chosen for that
edge. For example with $n=1$ see figure \vref{fig:Sample-invoice-edges}.
The relation does not need to be symmetrical, but when higher number
of closest neighbours will be used, the sets would have bigger overlap.
\item We can define a 'reading order of wordboxes'. In particular, based
on the idea that if two boxes do overlap in a projection to $y$ axis
by more than a given threshold, set to $50\,\%$ in our experiments,
they should be regarded to be in the same line for a human reader.
This not only defines an order of the boxes in which they will be
given as sequence to the network, but also assigns a line number and
order-in-line number to each box. To get more information, we can
run this algorithm again on a $90\text{\textdegree}$ rotated version
of the document. These integers are then subject to a positional embedding.
Note, that the exact ordering/reading direction (left to right and
top to bottom or vice versa) should not matter in the neural network
design, thus giving us the freedom to process any language.
\item Each box has 4 normalized coordinates (left, top, right, bottom) that
should be presented to the network also by positional embedding.
\end{itemize}
\item Textual:
\begin{itemize}
\item Each word can be presented using any fixed size representation, in
our case we will use tailored features common in other NLP tasks (e.g.
authorship attribution \cite{DBLP:journals/corr/abs-1208-3001}, named
entity recognition \cite{nadeau2007survey}, and sentiment analysis
\cite{Abbasi:2008:SAM:1361684.1361685}). The features per wordbox
are the counts of all characters, the counts of first two and last
two characters, length of a word, number of uppercase and lowercase
letters, number of text characters and number of digits. And finally,
if the word is in fact a number, then the number scaled and cropped
against different scales, zeroes for other text. The reason behind
these features is that in an invoice there would be a larger number
of named entities, ids and numbers, which are not easily embedded.
\item Trainable word features are employed as well, using convolutional
architecture over sequence of one hot encoded, deaccented, lowercase
characters (only alphabet, numeric characters and special characters
`` ,.-+:/\%?\$£€\#()\&'{}'', all others are discarded).
\end{itemize}
\item Image features:
\begin{itemize}
\item Each wordbox has its corresponding crop in the original PDF file,
where it is rendered using some font settings and also background,
which could be crucial to line-item table (or header) detection, if
it contains lines, for example, or different background color or gradient.
So the network receives a crop from the original image, offsetted
outwards to be bigger than the text area to see also the surroundings.
\end{itemize}
\end{itemize}
Each presented feature can be augmented, we have decided to do a random
$1\,\%$ percent perturbation on coordinates and textual features
representation.

\subsection{The architecture}

\begin{figure}[tb]
\includegraphics[width=1.1\columnwidth]{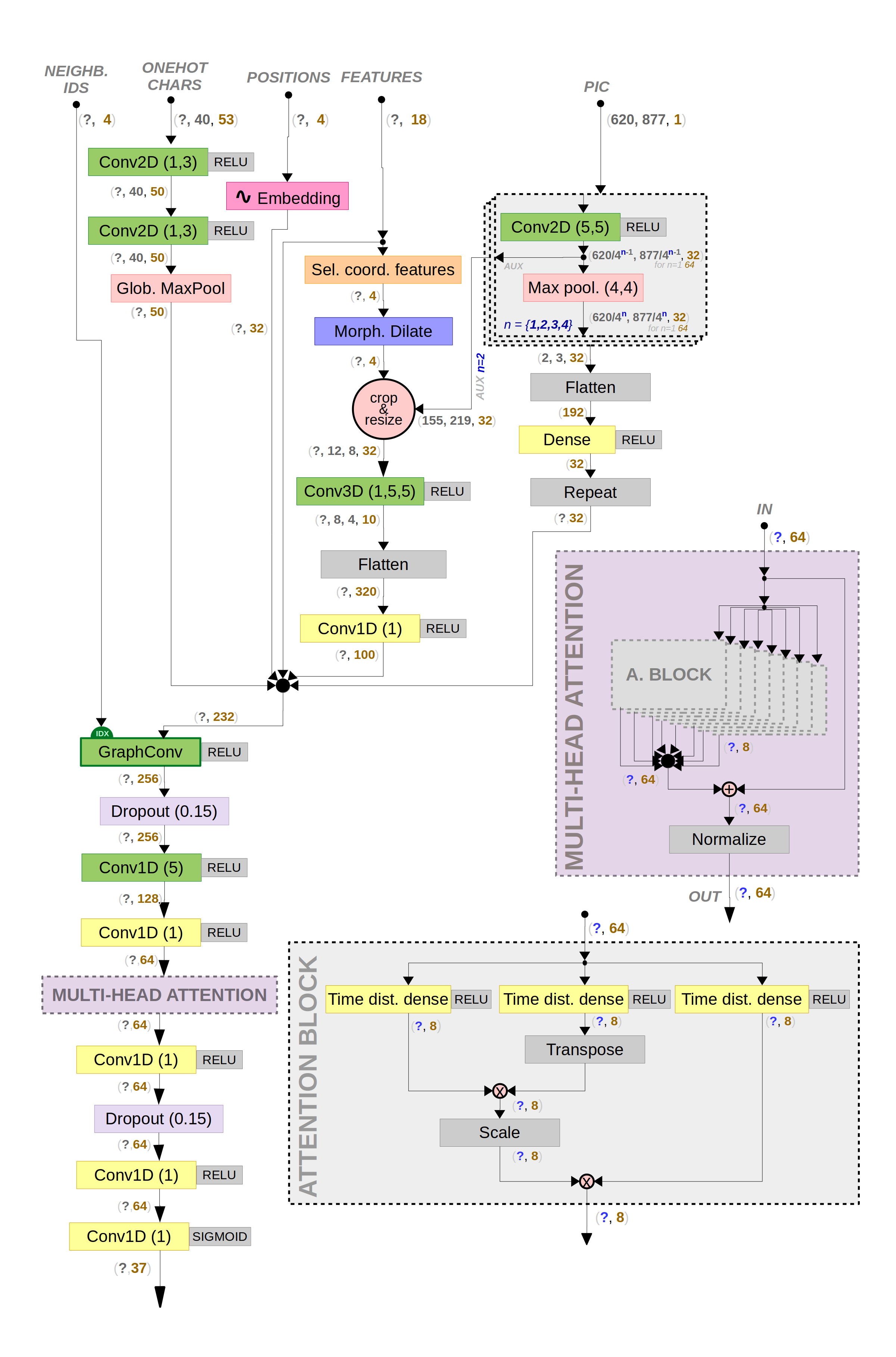}\caption{\label{fig:The-architecture}The model architecture. All features
are concatenated together before the self attention mechanism and
final layers. The cropping of the picture, embeddings and graph convolution
all happen inside the network. Note that $Conv1D(1)$ can be also
called a time distributed dense layer.}
\end{figure}

As can be seen in Figure \vref{fig:The-architecture} and as we have
stated before, the model uses 5 inputs - downsampled picture of the
whole document ($620\times877$), grayscaled; features of the wordboxes,
including their boundingbox coordinates; on-hot characters with 40
one-hot encoded characters per each wordbox; neighbour ids - integers
that define the neighbouring wordboxes on each side of the wordbox;
and finally integer positions of each field defined by the geometrical
ordering.

The positions are embedded by positional embeddings (defined and used
in \cite{2017arXiv170603762V,2018arXiv180703247L}, we use embedding
size equal to 4 dimensions for $sin$ and 4 for $cos$, with divisor
constant being $10000$) and then concatenated with other field features.

The picture is reduced by classical stacked convolution and maxpooling
approach and then from its inner representation, field coordinates
(left, top, right, bottom) are used to get a crop of a slightly bigger
area (using morphological dilation) which is then appended to the
field. Finally we have decided to give the model a grasp of the image
as whole - a connection to the whole image flattened and then processed
to 32 float features, which are also appended to each field's features.

Before attention, dense, or graph convolution layers are used, all
the input features are concatenated.

Our implementation of the graph convolution mechanism gathers features
from the neighbouring wordboxes, concatenates them and feeds into
a Dense layer. To note, our graph has a regularity that allows us
to simplify the graph convolution - there does exist an upper bound
on the number of edges for each node, so we do not need to use any
general form graph convolutions as in \cite{2016arXiv160505273N,2017arXiv170802218J}.

We have also employed a convolution layer over the ordering dimension
(called \emph{convolution over sequence} later in this text).

 The rest of the network handles images and crops. The final output
branch has an attention transformer module (from \cite{2017arXiv170603762V})
to be able to compare pairwise all the fields in hope that denser
and regular areas (of texts in a table grid) can be detected better.
Our attention transformer unit does not use causality, nor query masking
and has 64 units and 8 heads.

Finally, the output is a multilabel problem, so sigmoidal layers are
deployed together with binary crossentropy as the loss function.

The optimizer was chosen to be Adam. Model selected in each experimental
run was always the one that performed best on the validation set (of
the \emph{small} dataset) in terms of loss, while the \emph{patience}
constant was $10$ epochs. Batched data were padded by zeros per batch
(with zero sample weights). Class weights in our multi-task classification
problem were chosen based on positive class occurrences. The network
has $867k$ trainable parameters in total.

\section{Experiments}

The approach was tested on different data settings and different architectures.
There are 4 groups of experiments:
\begin{enumerate}
\item Comparing logistic regression baseline against the neural network.
\\
To note, logistic regression baselines use all the inputs except the
picture and trainable word embeddings. To inspect the importance of
neighbouring boxes, we have compared the baseline without neighbours
and the baseline with included information about one or more neighbours
at each side (if present).
\item The importance and effect of each block of layers and each input and
other parameters.\\
The choice of modules to test was 'convolution with dropout after
attention' to test the dropout layer, 'convolution over sequence'
for the importance of input ordering and attention. Experiments dropping
the graph convolution were done in variation of neighbours. Experiments
on anonymized dataset fall also into this category. We have also tested
the focal loss function \cite{2017arXiv170802002L}, note that we
do not vary final activations in our experiments here and use only
sigmoidal, because they had best performance in earlier development
process.
\item Specialization on a task where only line-items were classified and
specialization on a task with all but line-items.
\item Evaluating the model's adaptation performance on the big dataset (without
line-items).
\end{enumerate}
We will not be optimizing the number of neurons in the layers. The
training was done on a single GPU and ran approximately in $23$ epochs
for $5$ hours per experiment on small dataset.

\subsection{Results}

Table \vref{tab:Experiments-against-baseline} summarizes experiments
comparing the model against the logistic regression baseline, both
with varying number of neighbours (more than $2$ not shown, because
the results were not improving with the number of neighbours). The
logistic regression baselines did improve with more neighbours, but
failed to generalize. We can notice the big difference between line-item
table detection and other classes coming from a possible observation
that sometimes the table is the biggest one. The results also do reflect
the nature of a specialized structured document, which invoices indeed
are - to classify all the structured information is not easy for a
person not working with invoices.

On the other hand, the optimal number of neighbours for the final
architecture was 1, but we can notice, that 2 neighbours do help line-item
table body detections. We have designed the algorithm with more than
one neighbour in mind (with a single neighbour, the relation is not
symmetrical), so other positional features are possibly being exploited
more efficiently.

Table \vref{tab:Experiments-ablation} shows, that the multihead attention
module helps with generalization to unseen layouts, omitting the module
makes the network prioritize adaptation on already seen layouts. Also
without attention, the number of training epochs was twice (27) as
much as with attention (13). Focal loss, prioritizing rare classes,
does help line-item header detection, but is a cause for the decrease
of the nonline-item score micro metric, as rare classes contribute
less. 

The importance of the convolutional layer over the sequence might
come from our initial guess that this would give more importance to
beginnings and endings of lines of words. 

Table \vref{tab:Experiments-inputs} compares different inputs and
dataset choices. Although the architecture was optimized on the small
dataset, the results imply that the model has the capacity to adapt
and generalize also on bigger datasets. Looking at the anonymized
version of datasets, without some inputs, it can be concluded that
the network can learn to detect tight areas of evenly spaced words,
being the line-item table. Also even base text features help the model
generalize well. Overall the score on anonymized dataset means that
the positional information is passed correctly and embedded in a right
way for the network. 

In table \vref{tab:Experiments-less-targets} there can be seen that
the tasks of finding line-items and other structural information do
boost each other, with one exception being the header detection -
it does help adaptation, but when omitted, the generalization score
is higher.

The architecture provided on Figure \vref{fig:The-architecture} is
the '\textbf{complete model}', that uses binary crossentropy, all
inputs and all modules and a single neighbour at each side of each
box. Its generalization performance on unseen invoice types was $93\,\%$
on detecting line-items and $66\,\%$ for other classes ($87\,\%$
on similar layouts). To verify what the line-item detection scores
mean in practice, we have run the prediction on the sample invoices
(Figure \vref{fig:Sample-invoice-edges}), where our algorithm correctly
detected the line-item table up to 2 false positive words, which are
easily filtered out (heuristic filtering results are not reported).

\begin{table*}[t]
\caption{\label{tab:Experiments-against-baseline} }

\begin{tabular}{|c|c|c|c||c|c|c|}
\hline 
\multirow{3}{*}{Experiments against the baseline} & \multicolumn{3}{c||}{Adaptation} & \multicolumn{3}{c|}{Generalization}\tabularnewline
\cline{2-7} 
 & \multicolumn{2}{c|}{line-items} & others & \multicolumn{2}{c|}{line-items} & others\tabularnewline
\cline{2-7} 
 & body $F_{1}$ & header $F_{1}$ & micro $F_{1}$ & body $F_{1}$ & header $F_{1}$ & micro $F_{1}$\tabularnewline
\hline 
\hline 
complete model (without neighbours) & 0.9666 & 0.9969 & 0.8687 & 0.9242 & 0.9876 & 0.6609\tabularnewline
\hline 
\textbf{complete model (1 neighbour)} & 0.9738 & \textbf{0.9967} & \textbf{0.8790} & 0.9389 & \textbf{0.9864} & \textbf{0.6650}\tabularnewline
\hline 
complete model (with 2 neighbours)  & \textbf{0.9762} & 0.9963 & 0.8749 & \textbf{0.9408} & 0.9860 & 0.6629\tabularnewline
\hline 
logistic regression without neighbours  & 0.7594 & 0.9477 &  0.0004 & 0.7560 & 0.9362 & 0.0000\tabularnewline
\hline 
logistic regression with 1 neighbour  & 0.8664 & 0.9663 & 0.1482 & 0.8071 & 0.9461 & 0.0327\tabularnewline
\hline 
logistic regression with 2 neighbours  & 0.8939 & 0.9724 & 0.2276 & 0.8284 & 0.9493 & 0.0525\tabularnewline
\hline 
\end{tabular}
\end{table*}
\begin{table*}[t]
\caption{\label{tab:Experiments-ablation} }

\begin{tabular}{|c|c|c|c||c|c|c|}
\hline 
\multirow{3}{*}{Experiments with ablation} & \multicolumn{3}{c||}{Adaptation} & \multicolumn{3}{c|}{Generalization}\tabularnewline
\cline{2-7} 
 & \multicolumn{2}{c|}{line-items} & others & \multicolumn{2}{c|}{line-items} & others\tabularnewline
\cline{2-7} 
 & body $F_{1}$ & header $F_{1}$ & micro $F_{1}$ & body $F_{1}$ & header $F_{1}$ & micro $F_{1}$\tabularnewline
\hline 
\hline 
\textbf{complete model} & 0.9738 & 0.9967 & 0.8790 & \textbf{0.9389} & 0.9864 & \textbf{0.6650}\tabularnewline
\hline 
focal loss & 0.9735 & \textbf{0.9969} & 0.8557 & 0.9383 & \textbf{0.9878} & 0.6398\tabularnewline
\hline 
no convolution over sequence  & 0.9670 & 0.9945 & 0.8638 & 0.9101 & 0.9800 & 0.6237\tabularnewline
\hline 
no attention  & \textbf{0.9780} & 0.9967 & \textbf{0.8806} & 0.9348 & 0.9864 & 0.6487\tabularnewline
\hline 
no convolution with dropout after attention & 0.9646 & 0.9950 & 0.8435 & 0.9168 & 0.9807 & 0.6050\tabularnewline
\hline 
\end{tabular}
\end{table*}
\begin{table*}[t]
\caption{\label{tab:Experiments-inputs}}

\begin{tabular}{|c|c|c|c|c||c|c|c|}
\hline 
\multirow{3}{*}{Experiments with inputs variations} & \multirow{3}{*}{dataset} & \multicolumn{3}{c||}{Adaptation} & \multicolumn{3}{c|}{Generalization}\tabularnewline
\cline{3-8} 
 &  & \multicolumn{2}{c|}{line-items} & others & \multicolumn{2}{c|}{line-items} & others\tabularnewline
\cline{3-8} 
 &  & body $F_{1}$ & header $F_{1}$ & micro $F_{1}$ & body $F_{1}$ & header $F_{1}$ & micro $F_{1}$\tabularnewline
\hline 
\hline 
\textbf{complete model (all inputs)} & \textbf{small} & \textbf{0.9738} & \textbf{0.9967} & \textbf{0.8790} & \textbf{0.9389} & \textbf{0.9864} & \textbf{0.6650}\tabularnewline
\hline 
no text embeddings  & small & 0.9702 & 0.9921 & 0.7772 & 0.9108 & 0.9771 & 0.5118\tabularnewline
\hline 
no picture, only some text features  & anonym & 0.9694 & 0.9943 & 0.4518 & 0.9185 & 0.9805 & 0.4745\tabularnewline
\hline 
no picture, no text features  & anonym & 0.9588 & 0.9848 & 0.6836 & 0.8919 & 0.9549 & 0.2152\tabularnewline
\hline 
complete model (all inputs) & big & N/A & N/A & 0.8487 & N/A & N/A & N/A\tabularnewline
\hline 
\end{tabular}
\end{table*}
\begin{table*}[t]
\caption{\label{tab:Experiments-less-targets}}

\begin{tabular}{|c|c|c|c|c||c|c|c|}
\hline 
\multirow{3}{*}{Experiments with training target variations} & \multirow{3}{*}{dataset} & \multicolumn{3}{c||}{Adaptation} & \multicolumn{3}{c|}{Generalization}\tabularnewline
\cline{3-8} 
 &  & \multicolumn{2}{c|}{line-items} & others & \multicolumn{2}{c|}{line-items} & others\tabularnewline
\cline{3-8} 
 &  & body $F_{1}$ & header $F_{1}$ & micro $F_{1}$ & body $F_{1}$ & header $F_{1}$ & micro $F_{1}$\tabularnewline
\hline 
\hline 
\textbf{complete model (all outputs)} & \textbf{small} & \textbf{0.9738} & \textbf{0.9967} & \textbf{0.8790} & 0.9389 & \textbf{0.9864} & 0.6650\tabularnewline
\hline 
only line-items & small & 0.9027 & 0.9950  & N/A & 0.8762 & 0.9766 & N/A\tabularnewline
\hline 
no line-item header  & small & 0.9736 & N/A & 0.8777 & \textbf{0.9394} & N/A & \textbf{0.6731}\tabularnewline
\hline 
all but line-items & small & N/A & N/A & 0.8632 & N/A & N/A & 0.6247\tabularnewline
\hline 
complete model (other than line-items targets) & big & N/A & N/A & 0.8487 & N/A & N/A & N/A\tabularnewline
\hline 
\end{tabular}
\end{table*}

\section{Conclusions}

We have found a fully trainable method for table detection and content
understanding in structured documents, that is able to detect a specific
line-item table and extract only some information from other tables
even in the presence of imbalanced classes and multiple layouts, languages
and invoice types. Anonymized version of our dataset was published,
as no similar dataset has been publicly available to date.

Trying to detect line-item headers in a single model did lead the
model to underperform, with a hint to use focal loss for such task.
Also, we have discovered, that attention module was important to generalization
for new invoice types, while using only close neighbours did lead
to better adaptation on already seen layouts. 

The system's ability to correctly scale to completely new invoice
types is successfully verified for the line-item table detection task
at 93\% and measured to be 66\% on 35 'other' classes.

Future work can include line-item table extractions, architecture
and hyperparameter tuning for bigger datasets, experiments with the
usage of different text features or embeddings and image augmentations.
It could be also measured how many annotations are needed for the
'other' classes to adapt onto new invoice types.

\appendices{}

\section*{Acknowlegment}

The work was supported by the grant \ifanonym
\else SVV-2017-260455. \fi We would also like to thank to the annotation
team and the rest of the research team.

\bibliographystyle{IEEEtran}
\bibliography{LineItemAndTableUnderstandingInStructuredDocuments}
\begin{IEEEbiography}[{\fbox{\begin{minipage}[t][1.25in]{1in}%
Replace this box by an image with a width of 1\,in and a height of
1.25\,in!%
\end{minipage}}}]{Your Name}
 
\end{IEEEbiography}

\begin{IEEEbiographynophoto}{Coauthor}
Same again for the co-author, but without photo
\end{IEEEbiographynophoto}

\end{document}